 \documentclass[prl, twocolumn, showpacs, amsmath]{revtex4}

\usepackage{dcolumn}% Align table columns on decimal point
\usepackage{bm}% bold math
\usepackage{graphicx}% Include figure files
\usepackage{color}
\usepackage{subfigure}
\usepackage{hyperref}
\usepackage{latexsym}
\usepackage{amsthm}
\usepackage{amssymb}
\usepackage{multirow}
\usepackage[english]{babel}
\usepackage{comment}

\begin{document}

\newcommand{\filippo}[1]{{\bf\color{red}#1}}
\newcommand{\gin}[1]{{\bf\color{blue}#1}}
\newcommand{\avg}[1]{\langle{#1}\rangle}
\newcommand{\Avg}[1]{\left\langle{#1}\right\rangle}

\title{Percolation in
real multiplex networks}

\author{Ginestra Bianconi}
\affiliation{School of Mathematical Sciences, Queen Mary University of
  London, London, E1 4NS, United Kingdom}
\email{ginestra.bianconi@gmail.com}
\author{ Filippo Radicchi}
\affiliation{Center for Complex Networks and Systems Research, School
  of Informatics and Computing, Indiana University, Bloomington, IN
  47408, USA}
  \email{filiradi@indiana.edu}

\begin{abstract}
We present an exact mathematical
framework able to describe site-percolation transitions in real multiplex networks.
Specifically, we consider the average percolation diagram
valid over an infinite number of random configurations
where nodes are present in the system with 
given probability. The approach relies on the locally treelike
ansatz, so that it is expected to accurately reproduce
the true percolation diagram of sparse multiplex networks
with negligible number of short loops. 
%Performances of the new framework are tested 
 %in social, biological, and transportation 
%multiplex graphs. 
The performance of our theory is tested 
 in social, biological, and transportation 
multiplex graphs. 
When compared 
against previously introduced 
methods, we observe improvements in the prediction
of the percolation diagrams in all networks analyzed. 
Results from our method confirm previous
claims about the robustness of real multiplex networks, in the sense that 
the average connectedness of the system  does 
not exhibit any significant abrupt change
as its individual components are randomly destroyed.
\end{abstract}


\maketitle
Many, if not all, real-world networks are coupled with or
interact with other networks~\cite{buldyrev2010catastrophic}. The notion 
of multiplex network represents a way 
of accounting for such
fundamental feature~\cite{boccaletti2014structure, kivela2014multilayer}. 
Loosely speaking, a multiplex
is defined as a network composed 
of $N$ nodes connected
in some way through a set of edges 
that can assume $M$ possible colors or flavors. 
Often, it is convenient to think of the system as
a layered network, where individual network 
layers are generated by grouping together edges 
with the same color.  
The representation of a real system as a multiplex network 
is appropriate  in disparate contexts, such
as (but not limited to) 
social networks sharing the same
actors~\cite{szell2010multirelational, mucha2010community}, 
multimodal transportation graphs sharing 
common geographical
locations~\cite{barthelemy2011spatial,cardillo2013emergence}, and coupled
 networks of power distribution
and communications~\cite{buldyrev2010catastrophic}.

The first, and probably the most important, 
model studied on multiplex
networks, is the so-called
site-percolation 
model~\cite{stauffer1991introduction, buldyrev2010catastrophic}.
 This model
serves as a proxy to 
quantify the robustness of
networked systems under random failures, by
monitoring how the connectedness
at the macroscopic level changes
as a function of the amount of microscopic 
damages of individual nodes~\cite{albert2000error, cohen2000resilience,
  callaway2000network, newman2010networks}.
In their seminal paper, Buldyrev {\it et al}. showed that 
multiplex networks composed of random network models
with negligible overlap
undergo a discontinuous percolation 
transitions when interdependencies are introduced between the nodes in different layers~\cite{buldyrev2010catastrophic}.
The model has been studied extensively
on ensembles of multiplex networks~\cite{son2012percolation}, where it has been found that the transition is not only discontinuous but also  hybrid, i.e., it displays a square root singularity~\cite{baxter2012avalanche}.
This theory has been extended in different directions to correlated
multiplex networks and more general multilayer 
structures~\cite{parshani2010interdependent,min2014network,bianconi2014multiple,cellai2016multiplex}.
Among the different types of correlations that can be found  in
multiplexes, link overlap~\cite{bianconi2013statistical} plays a 
major role because of its ubiquity in real network structures~\cite{szell2010multirelational,cardillo2013emergence}.
Despite some earlier works on duplex
networks~\cite{hu2013percolation}, 
percolation theory  in presence of link overlap has been elusive
until recently. 
An appropriate
mathematical framework able to describe the emergence of the
giant component in arbitrary multiplex networks has been introduced in 
Refs.~\cite{baxter2016correlated,PhysRevE.94.032301} to characterize the percolation transition in ensemble of multiplex networks~\cite{bianconi2013statistical}. In these papers, it has been found that in multiplex networks the percolation transition is always discontinuous with the only exception of the trivial case in which all the layers completely overlap.

Much less attention has been devoted to the analysis of the
percolation model on real-world multiplexes.
These systems generally exhibit overlap only in a 
small fraction of core edges that are able to
keep the system connected without leading to any significant abrupt
transition~\cite{radicchi2015percolation}. The result of Ref.~\cite{radicchi2015percolation}
has been obtained through the development of a mathematical 
framework able to approximate the percolation diagram
of arbitrary multiplexes. However in the method of Ref.~\cite{radicchi2015percolation},
a good approximation of the true percolation diagram is
granted only if the network obtained from the
overlap of the layers is either fragmented in vanishing clusters, or
it contains a unique giant component~\cite{PhysRevE.94.032301}.
In the intermediate case when multiple nonvanishing clusters
are present in the overlap network, the method developed in~\cite{radicchi2015percolation},
as well as those used in~\cite{cellai2013percolation}, describes a different type of
model not compatible with the one of the site-percolation model~\cite{min2015link,baxter2016correlated,PhysRevE.94.032301}.  
The goal of this letter is to
introduce an exact mathematical theory  able to provide the
solution of the percolation model in arbitrary multiplex
networks. In this approach, we take as input the topology of the
multiplex to draw  the entire percolation diagram. 
Such a diagram
approximates how the relative size of the largest mutually connected cluster
in the graph varies as a function of the microscopic probability of
individual nodes to be present in the system.

The present theory  is developed for multiplexes with
arbitrary number of layers.
The only approximation used is the so-called locally treelike ansatz,
according to which nearest-neighbors of every node are not connected
among themselves~\cite{dorogovtsev2008critical}. We remark that this
approximation may be not justified in many real
systems~\cite{PhysRevE.93.030302}. On the other hand,
all theoretical approaches generated so far in this context suffer from the same exact
limitation, including methods deployed for the description of
the (simpler) percolation model in isolated
networks~\cite{PhysRevLett.113.208701, PhysRevLett.113.208702}.
Whereas in the context of isolated networks improved methods
exist~\cite{PhysRevE.93.030302}, 
corrections to frameworks valid for 
multiplex networks do not seem 
as straightforward.

For illustrative purposes, we will consider here 
only the case
of a multiplex composed of $M=2$ layers.
The 
general case 
$M \geq 2$ is presented in the SM. 
Without loss of generality, we assume
that a multiplex network $G$ composed
of $N$ nodes is given. 
Every node $i \in G$ appears in both
layers so that the  failure of a node in one layer 
implies the simultaneous 
failure of its copy in the other layer.  
Connections among pairs of nodes are specified 
in the adjacency matrices of the layers: 
on each individual  layer $\alpha = 1, 2$,
a connection between the nodes $i$ and $j$ exists
if $a_{ij}^{[\alpha]} =  a_{ji}^{[\alpha]}  = 1$, whereas no connection 
between nodes $i$ and $j$ 
exists in layer $\alpha$ if
$a_{ij}^{[\alpha]} =  a_{ji}^{[\alpha]} = 0$.
For convenience of notation, we define for every pair of nodes
$i$ and $j$ the multilink vector \cite{bianconi2013statistical} $\vec{m}_{ij} = \left( a_{ij}^{[1]},
  a_{ij}^{[2]}\right)$, so that the entire topological information
of the multiplex is stored in $N(N-1)/2$ two-dimensional vectors.
This represents the input of the 
mathematical framework that we are going to describe below.

We consider the ordinary version of the
site-percolation model on multiplex networks, where
every node is present in the system 
with probability $p$~\cite{stauffer1991introduction}.
Nodes that are present form clusters of connected nodes.
Depending on the value of $p$, nodes may be or may not 
form a
mutually connected giant 
component (MCGC)~\cite{buldyrev2010catastrophic}. 
The MCGC is  identified in a recursive manner and is  composed 
by all the  vertices that  are connected by at least by one path (internal to the MCGC) in each layer.
In infinitely large networks, the MCGC exists for values of
$p>p_c$, whereas it doesn't exists
if $p \leq p_c$. Further, with the exception of
the trivial case of duplex networks whose layers completely overlap,
the MCGC emerges discontinuously~\cite{baxter2016correlated,PhysRevE.94.032301,bianconi2013statistical}.
In finite systems, such as real-world multiplexes, although the
transition is not properly defined, we can still monitor the behavior
of the MCGC as a function of the probability $p$, and define 
a pseudo-transition point $p_c$. Such a threshold represents a
good proxy to measure how robust is a given multiplex, as it
indicates the fraction of nodes that must be in a functional state
in order to preserve a macroscopic connectedness in the system.
Additional information about system robustness can be 
gauged from the entity of the variation of the MCGC around this point.
Whereas the latter is generally difficult to measure from a finite
number of numerical
simulations, it can be instead easily derived from an
analytic framework, such as the one described below, 
that is able to well describe average values of the MCGC
over an infinite number of realizations of the percolation model.

%\gin{Here  the section in which I made the modifications starts.}

The mathematical framework that
allows us to compute how the size of the MCGC varies as a function of
the microscopic probability $p$ consists in a 
set of self-consistent messages exchanged by
pairs of connected nodes~\cite{mezard2009information}.
A similar  message-passing algorithm is a well-established method to
detect the giant component in single
networks~\cite{newman2010networks}. In ordinary percolation on single
networks, the message between 
node $i$ and node $j$  indicates the probability that node $i$ connects node $j$ to the giant component. In our multiplex percolation problem, instead,  the  message between node $i$ and node $j$ includes the information about the specific set of layers where node $j$ is connected to the MCGC.

A  message can be delivered from 
node $i$ to node $j$ only if a connection between 
node $i$ and node $j$ exists
in the system, i.e. if $\vec{m}_{ij}\neq \vec{0}$. 
 Please note that, whereas the network is undirected,
messages instead travel in the system following specific directions,
so that a message proceeding in the direction $i \to j$ is not necessarily 
identical to the message travelling in the opposite direction 
$j \to i$.

Let us define a vector $\vec{n} = (n^{[1]}, n^{[2]})$ of elements $n^{[\alpha]}=0,1$ and let us consider a   pair of nodes $i$ and $j$ connected by a multilink $\vec{m}_{ij}\neq \vec{0}$.
 The  message $s_{i \to j}^{\vec{m}_{ij} ,
  \vec{n}}$ indicates the probability that node $i$ connects node $j$ to the MCGC in all the layers $\alpha$ where $n^{[\alpha]}=1$.
 For example, given two nodes $i$, and $j$ connected by a $\vec{m}_{ij}=(1,1)$, $s_{i \to j}^{(1,1), (1,1)}$ indicates the probability that node $i$ connects node $j$ to the MCGC  in  both
layers. Similar straightforward definitions are valid for the other  messages.
Out of all the possible messages $s_{i \to j}^{\vec{m}_{ij} ,
  \vec{n}}$, there is a  set of  trivial messages that are always equal to zero.
  In fact   node $i$  cannot connect node $j$ to the MCGC in a layer $\alpha$ if the two nodes are not connected in that layer. Therefore if  $m^{[\alpha]}_{ij}=0$ we cannot have $n^{[\alpha]}=1$. 
It follows that   $
s_{i \to j}^{(1,0), (0,1)} = s_{i \to j}^{(1,0), (1,1)} 
= s_{i \to j}^{(0,1), (1,0)} = s_{i \to j}^{(0,1), (1,1)} =
0$ or equivalently   $s_{i\to j}^{\vec{m}_{ij}\vec{n}}=0$,  if 
$n^{[1]}(1-m^{[1]}_{ij})+n^{[2]}(1-m^{[2]}_{ij})\neq 0$.

Furthermore, we can omit the separate treatment of the messages $s_{i \to j}^{\vec{m}_{ij}, (0,0)}$ since we always have the normalization condition $s_{i \to j}^{\vec{m}_{ij}, (0,0)} = 1-s_{i \to j}^{\vec{m}_{ij}, (0,1)}-s_{i \to j}^{\vec{m}_{ij}, (1,0)}-s_{i \to j}^{\vec{m}_{ij}, (1,1)}$.
The remaining  five messages $s_{i \to
  j}^{(1,1), (1,1)}$, $s_{i \to j}^{(1,1), (1,0)}$, $s_{i \to
  j}^{(1,1), (0,1)}$, $s_{i \to j}^{(1,0), (1,0)}$, and $s_{i \to
  j}^{(0,1), (0,1)}$ obey the following system
of coupled nonlinear equations
\begin{equation}
\begin{array}{l}
s_{i \to j}^{(1,1), (1,1)} = s_{i \to j}^{(1,0), (1,0)}  = s_{i \to
  j}^{(0,1), (0,1)}  = 
\\p \, \left[1 - \prod_{\ell \in N(i)
    \setminus j} (1 - z_{\ell \to i}^{[1]}) -  \prod_{\ell \in N(i)
    \setminus j} (1 - z_{\ell \to i}^{[2]}) \right.
\\
\left.
+ \prod_{\ell \in N(i)
    \setminus j} (1 - z_{\ell \to i}^{[1,2]})  \right] 
\end{array}
\;, 
\label{eq:1}
\end{equation}
\begin{equation}
s_{i \to j}^{(1,1), (1,0)} = p \, \left[\prod_{\ell \in N(i)
    \setminus j} (1 - z_{\ell \to i}^{[2]}) -  \prod_{\ell \in N(i)
    \setminus j} (1 - z_{\ell \to i}^{[1,2]})  \right] \; ,
\label{eq:2}
\end{equation}
and
\begin{equation}
s_{i \to j}^{(1,1), (0,1)} = p \, \left[\prod_{\ell \in N(i)
    \setminus j} (1 - z_{\ell \to i}^{[1]}) -  \prod_{\ell \in N(i)
    \setminus j} (1 - z_{\ell \to i}^{[1,2]})  \right] \; .
\label{eq:3}
\end{equation}
In the previous equations, we have indicated with $ N(i)$ the neighbors of node $i$, i.e. 
$N(i) = \{ j \in G \, |\,  \vec{m}_{ij}\neq \vec{0} \}$ and 
we have defined  
\begin{equation}
z^{[1]}_{i \to j} = s_{i \to j}^{\vec{m}_{ij}, (1, 0)} + s_{i \to j}^{\vec{m}_{ij}, (1, 1)}
\; , 
\label{eq:z1}
\end{equation}
\begin{equation}
z^{[2]}_{i \to j} = s_{i \to j}^{\vec{m}_{ij}, (0, 1)} + s_{i \to j}^{\vec{m}_{ij}, (1, 1)} \; ,
\label{eq:z2}
\end{equation}
and 
\begin{equation}
z^{[1,2]}_{i \to j} = s_{i \to j}^{\vec{m}_{ij}, (0, 1)} + s_{i \to
  j}^{\vec{m}_{ij}, (1, 0)} + s_{i \to j}^{\vec{m}_{ij}, (1, 1)} 
%= z^{[1]}_{i \to j} + z^{[2]}_{i \to j} -  s_{i \to j}^{\vec{m}_{ij},
%(1, 1)} 
\; .
\label{eq:z3}
\end{equation}
Here, $z_{i \to j}^{[1]}$ represents the total probability node $i$ connects node $j$ to the MCGC through links of layer $\alpha = 1$;
$z_{i \to j}^{[2]}$ is the same as $z_{i \to j}^{[1]}$, but for layer
$\alpha =2$; 
$z_{\ell \to i}^{[1,2]}$ equals instead  the probability
that node $i$ connects node $j$ to the MCGC at least in one layer.
Eqs.~(\ref{eq:1}),~(\ref{eq:2}),
and ~(\ref{eq:3}) connect in a self-consistent manner 
the various messages, accounting
for the presence of edge overlap among layers.
We remark also that the topology of the 
network is given, so that
only the non-trivial  messages appearing in the
Eqs.~(\ref{eq:1}),~(\ref{eq:2}),
and ~(\ref{eq:3}) are actually non-zero. 
%or equivalently, if  $n^{[\alpha]}=1$ for some $\alpha$ such that  $m^{[\alpha]}_{ij}=0$.
%For instance, if the multilink
%$\vec{m}_{ij} = (1,1)$, so that the edge $(i,j)$ is present
%in both layers, then the messages that we have to consider are  
%$s_{i \to j}^{(1,1), (1,1)}$, $s_{i \to j}^{(1,1), (1,0)}$, and $s_{i
 % \to j}^{(1,1), (0,1)}$. 
%If instead $\vec{m}_{ij} = (1,0)$, so that $i$ and $j$ are connected
%only
%on layer $1$, then we need to consider only the message 
%$s_{i \to j}^{(1,0), (1,0)}$. Similarly, if $\vec{m}_{ij} = (0,1)$,
%then the message to consider is $s_{i \to j}^{(0,1), (0,1)}$.
From Eq.~(\ref{eq:1}), we note that $s_{i \to j}^{(1,1), (1,1)}$, 
$s_{i \to j}^{(1,0), (1,0)}$, $s_{i \to j}^{(0,1), (0,1)}$ are
determined as the probability that node $i$ is present, thus the
factor $p$, multiplied by the probability that node $i$ is receiving
(or not receiving)
coherent messages in both layers.  
The message $s_{i \to  j}^{(1,1), (1,0)}$ defined in
Eq.~(\ref{eq:2}) is computed from the messages incoming from neighboring nodes different from $j$. Its value is given by   the probability that the node $i$ is
present multiplied by the probability that node $i$ is connected to
the MCGC in layer $\alpha=1$, 
but is not connected to the MCGC in layer $\alpha = 2$. The message $s_{i \to
  j}^{(1,1), (0,1)}$ of Eq.~(\ref{eq:3}) is defined in analogous
manner. We note two fundamental things common in 
the r.h.s. of Eqs.~(\ref{eq:1}),~(\ref{eq:2}),
and ~(\ref{eq:3}): (i) Probabilities are estimated under the locally
treelike approximation, hence the appearance of products
of probabilities for (hypothetically) nonconnected neighbors; (ii)
When calculating 
message for the pair $i \to j$, we always exclude contributions of node $j$
in the products, thus avoiding for the presence of immediate
backtracking messages. The inclusion of the messages
$s_{i \to j}^{(1,1), (0,1)}$ and $s_{i \to j}^{(1,1), (1,0)}$
represent the fundamental difference between the 
current method and
the one developed in Ref.~\cite{radicchi2015percolation}.
These terms serve to account for the possibility that
the overlap graph may be divided in different clusters connected by  distant single layer links. 
In fact, these messages, by  preserving the information about the single layers connected to the MCGC,  allow the algorithm to propagate  from  cluster to cluster~\cite{PhysRevE.94.032301}.
For a given value of $p$, Eqs.~(\ref{eq:1}),~(\ref{eq:2}),
and ~(\ref{eq:3}) can be solved by iteration. 
The solutions of these equations are then plugged into
\begin{equation}
\begin{array}{l}
r_i = p \left[  1 - \prod_{j \in N(i)} (1 - z^{[1]}_{j \to i})  \right.
\\
\left. - \prod_{j \in N(i)} (1 - z^{[2]}_{j \to i}) + \prod_{j \in
  N(i)} (1 - z^{[1,2]}_{j \to i})\right]
\end{array} 
\label{eq:node}
\end{equation}
to estimate the probability $r_i$ that node $i$ belongs to
the MCGC. Finally, the average size of the MCGC is calculated as
\begin{equation}
{P}^{\textrm{(th)}}_\infty = \frac{1}{N} \, \sum_{i=1}^N \, r_i \; .
\label{eq:perc}
\end{equation}
By changing the value of $p \in [0,1]$ and solving Eqs.~(\ref{eq:1})-(\ref{eq:perc}), one can draw the entire
percolation diagram for a given multiplex.
%\gin{Here the section in which I made my modifications finishes.}

%%%%%%%%%
\begin{table*}[!htb]
\begin{center}
\begin{tabular}{|l|l|r|r|r|r|r|r|r|r|r|r|r|r|}
\hline 
Network 
& Layers &$N$ & $E^{[1,2]}$ & $E^{[1]}$ & $E^{[2]}$ & $O$ &
                                                            ${p}^{\textrm{(num)}}_c$ &
                                                                    ${p}^{\textrm{(Rad)}}_c$
& ${\hat{P}}^{\textrm{(Rad)}}_\infty$ & $\epsilon^{\textrm{(Rad)}}$ &
                                                                     ${p}^{\textrm{(th)}}_c$
  & ${\hat{P}}^{\textrm{(th)}}_\infty$ & $\epsilon^{\textrm{(th)}}$ 
\\
\hline 
%\multirow{3}{*}{\rotatebox[origin=c]{90}{\parbox[c]{1cm}{\centering US 
%  Air}}} & 
\multirow{3}{*}{\parbox{2.8cm}{{\it US 
  Air Transportation}~\cite{radicchi2015percolation}}} & 
{\it Am. Air.} -- {\it Delta} &
$84$ & $136$ & $380$ & $748$ & $0.11$ & $0.29$ & $0.24$ & $0.03$ & $0.01$ & $0.17$ & $0.01$ & $0.01$
\\
& {\it Am. Air.} -- {\it United} &
$73$ & $136$ & $322$ & $404$ & $0.16$ & $0.30$ & $0.20$ & $0.00$ & $0.01$ & $0.15$ & $0.00$ & $0.01$
\\ 
& {\it Delta} -- {\it United} &
$82$ & $112$ & $696$ & $452$ & $0.09$ & $0.27$ & $0.26$ & $0.03$ & $0.03$ & $0.17$ & $0.02$ & $0.01$
\\ \hline

\multirow{3}{*}{\parbox{2.8cm}{
 {\it Caenorhabditis Elegans}~\cite{chen2006wiring, de2014muxviz}}} & 
%1 -- 2 &
{\it Electric} -- {\it Chem.~Mon.} &
$238$ & $222$ & $748$ & $1,324$ & $0.10$ & $0.45$ & $0.26$ & $0.00$ & $0.01$ & $0.22$ & $0.00$ & $0.02$
\\ 
%& 1 -- 3 
& {\it Electric} -- {\it Chem.~Pol.} &
$252$ & $324$ & $698$ & $2,586$ & $0.09$ & $0.36$ & $0.23$ & $0.00$ & $0.02$ & $0.20$ & $0.00$ & $0.02$
\\
%& 2 -- 3 
& {\it Chem.~Mon.} -- {\it Chem.~Pol.} &
$259$ & $1,260$ & $514$ & $1,892$ & $0.34$ & $0.22$ & $0.11$ & $0.00$ & $0.01$ & $0.10$ & $0.00$ & $0.01$
\\ \hline

\multirow{3}{*}{\parbox{2.8cm}{{\it Drosophila Melanogaster}~\cite{stark2006biogrid, de2015structural}}} & 
%1 -- 2 &
{\it Direct} -- {\it Supp.~Gen.} &
$676$ & $132$ & $1,204$ & $2,556$ & $0.03$ & $0.67$ & $0.68$ & $0.05$ & $0.01$ & $0.60$ & $0.03$ & $0.01$
\\ 

%& 1 -- 3 &
& {\it Direct} -- {\it Add.~Gen.} &
$625$ & $98$ & $948$ & $1,950$ & $0.03$ & $0.75$ & $0.85$ & $0.08$ & $0.02$ & $0.75$ & $0.00$ & $0.01$
\\

%& 2 -- 3 &
& {\it Supp.~Gen.} -- {\it Add.~Gen.} &
$557$ & $936$ & $1,906$ & $1392$ & $0.22$ & $0.26$ & $0.17$ & $0.00$ & $0.01$ & $0.14$ & $0.00$ & $0.01$
\\ \hline 

\multirow{3}{*}{\parbox{2.8cm}{
 {\it Homo Sapiens}~\cite{stark2006biogrid, de2014muxviz}}} & 
%1 -- 2 &
{\it Direct} -- {\it Physical} &
$9,553$ & $23,930$ & $60,824$ & $112,440$ & $0.12$ & $0.42$ & $0.04$ & $0.00$ & $0.00$ & $0.04$ & $0.00$ & $0.00$
\\
%& 1 -- 3 &
& {\it Direct} -- {\it Supp.~Gen.} &
$4,465$ & $2,724$ & $36,658$ & $26,742$ & $0.04$ & $0.23$ & $0.18$ & $0.00$ & $0.00$ & $0.16$ & $0.00$ & $0.00$
\\
%& 2 -- 3 &
& {\it Physical} -- {\it Supp.~Gen.} &
$5,202$ & $4,436$ & $80,560$ & $30,754$ & $0.04$ & $0.48$ & $0.09$ & $0.00$ & $0.00$ & $0.08$ & $0.00$ & $0.00$
\\ \hline 

\multirow{3}{*}{\parbox{2.8cm}{
  {\it NetSci Co-authorship}~\cite{de2015identifying}}} &
{\it {\tiny physics.data-an}} -- {\it  {\tiny cond-mat.dis-nn}} & 
$1,400$ & $5,112$ & $2,278$ & $1,208$ & $0.59$ & $0.32$ & $0.09$ & $0.00$ & $0.07$ & $0.09$ & $0.00$ & $0.08$
\\
& {\it  {\tiny physics.data-an}} -- {\it  {\tiny cond-mat.stat-mech}} & 
$709$ & $2,318$ & $896$ & $244$ & $0.67$ & $0.62$ & $0.10$ & $0.00$ & $0.14$ & $0.10$ & $0.00$ & $0.15$
\\
& {\it  {\tiny cond-mat.dis-nn}} -- {\it  {\tiny cond-mat.stat-mech}} & 
$499$ & $1,004$ & $530$ & $322$ & $0.54$ & $0.86$ & $0.19$ & $0.00$ & $0.12$ & $0.19$ & $0.00$ & $0.13$
\\ \hline 

\end{tabular}
\end{center}
\caption{List of real-world multiplexes analyzed. The first column 
  identifies the name of the system analyzed, and the reference(s) of the 
paper(s) where such a system has been previously considered. In the second column,
we report the names of the different pairs of layers used to construct 
duplex networks. For each of them, we report in the following columns:
number of nodes ($N$), twice the number of edges shared by both layers 
($E^{[1,2]}$), twice the number of edges present only in the first or the second 
layer ($E^{[1]}$ and $E^{[2]}$), normalized overlap among the layers [$O =
E^{[1,2]} / (E^{[1,2]} + E^{[1]} + E^{[2]})$], best estimate of the 
percolation threshold ($p^{\textrm{(num)}}_c$), predictions according to the method
of Ref.~\cite{radicchi2015percolation} for the threshold and height of 
the jump of the transition [${p}^{\textrm{(Rad)}}_c$
and ${\hat{P}}^{\textrm{(Rad)}}_\infty$],  value of the error
$\epsilon^{\textrm{(Rad)}}$ with respect to the numerical curve,
predictions according to 
the current framework 
for the threshold and height of 
the jump of the transition [${p}^{\textrm{(th)}}_c$
and ${\hat{P}}^{\textrm{(th)}}_\infty$], and value of the error
$\epsilon^{\textrm{(th)}}$ with respect to the numerical curve.
Numerical values in the rightmost columns of the table contain up to
two significant
digits, therefore $0.00$ stands for values smaller than $0.01$.}
\label{table}
\end{table*}
%%%%%%%%%

%%%%%%%%%%%
\begin{figure*}[!htb]
\includegraphics[width=0.9\textwidth]{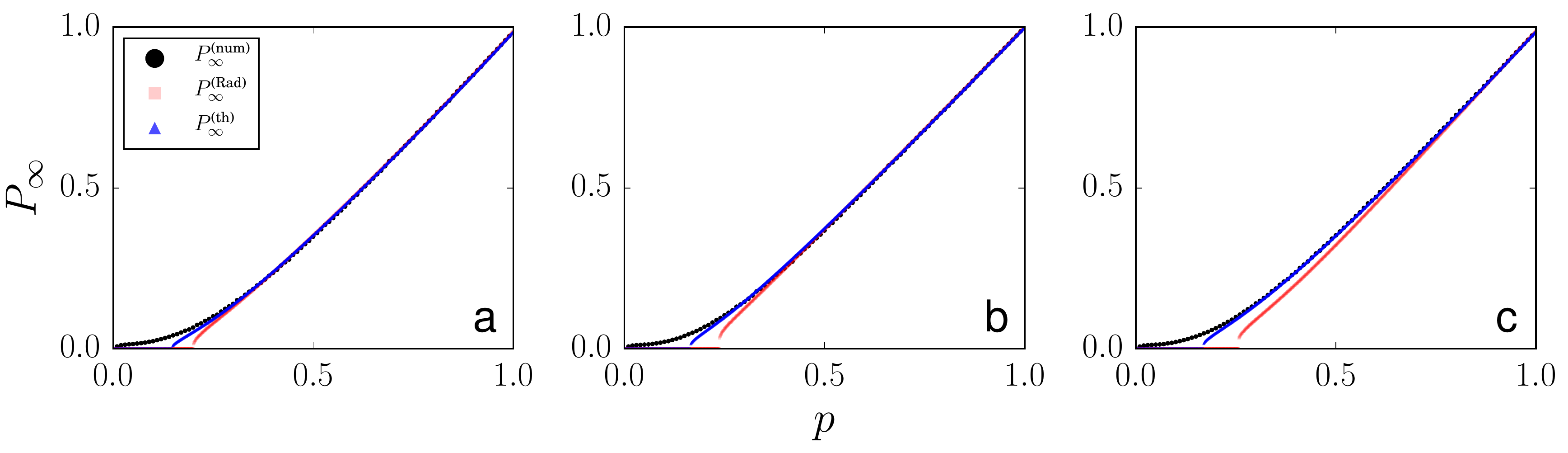}
\caption{Percolation diagram for US air transportation duplexes. (a) The
  system is obtained by combining American Airlines and Delta routes. 
We consider only US domestic flights operated in January, 2014, and
construct the duplex network where airports are nodes, and connections
on the layers are determined by 
the existence of at least a flight between the two locations. 
In the percolation diagram, black circles are results of numerical
simulations, red squares are results from the framework of
Ref.~\cite{radicchi2015percolation}, and
blue triangles are results obtained from the current method. 
%Vertical lines identify the critical threshold obtained with the
%various methods: $p^{\textrm{(num)}}_c$ (full black line), 
%$p^{\textrm{(Rad)}}_c$ (dotted red line), $p^{\textrm{(th)}}_c$
%(dashed blue line).  
b) Same as in a, but for the combination 
of American Airlines and United flights. c) Same as in a, but for the
combination of Delta and 
United flights.}
\label{figure}
\end{figure*}
%%%%%%%%%%%

To test the performance of the theory, we consider $15$ real-world
multiplexes (see Table~\ref{table} for the list of networks). We
compare the numerical solutions of our method with the solution of
the framework of Ref.~\cite{radicchi2015percolation}.  
For shortness, we indicate with $P^{\textrm{(Rad)}}_\infty$
the order parameter computed 
according to  Ref ~\cite{radicchi2015percolation}. 
Critical thresholds according to both approximations are obtained with a binary search strategy
able to identify the value of $p$ where the order parameter
$P_\infty$ changes from zero to a value larger than zero. We indicate
with $p^{\textrm{(th)}}_c$ the threshold obtained with the current framework,
and with $p^{\textrm{(Rad)}}_c$ the one 
computed with the method of Ref ~\cite{radicchi2015percolation}.
Further, we use as a term of comparison the
ground truth obtained through numerical simulations of the
percolation model. Values of the order parameter $P^{\textrm{(num)}}_\infty$ 
are obtained by averaging over
$10,000$ random configurations
of the percolation model for a given value of the probability $p$.
For numerical simulations, the critical threshold
$p^{\textrm{(num)}}_c$
is estimated as the value of $p$ where the
susceptibility reaches its maximum~\cite{radicchi2014predicting}.
We stress that the value of $p_c$ obtained from numerical simulations
characterizes only the average behaviour of the multiplex network
under random damage and that the  position of the transition for a
given realization of the initial  damage might have  large
fluctuations  for multiplex networks of small size.  
Further, we measure the overall performance of the theoretical
approaches to approximate the percolation phase diagram
obtained from numerical simulations using the distance measure~\cite{Melnik11}
\begin{equation}
\epsilon^{\text(x)} = \int_{0}^{1} \, | P_{\infty}^{\text{(x)}} (p) -
P_{\infty}^{\text{(num)}} (p) | \; dp \; ,
\label{eq:err}
\end{equation}
with $x = \textrm{Rad}$, or $x = \textrm{th}$.

In Fig.~\ref{figure}, we show the percolation diagram
of multiplexes representative for the air transportation network
within the US~\cite{radicchi2015percolation}. 
Our framework
provides better prediction of the true phase diagram than the
method developed in Ref.~\cite{radicchi2015percolation}.
Improvements are apparent from the fact that
the predicted curve is always closer to the true one.
This is demonstrated from the fact that $\epsilon^{\text(th)} \leq
\epsilon^{\text(Rad)}$ (Table~\ref{table}).
The same qualitative result is also visible in the other networks
analyzed (see SM). Overall,
we note that the framework of
Ref.~\cite{radicchi2015percolation} generates results
almost identical to those of the method proposed here
(the only clear exception found is the multiplex representing
interactions among genes and proteins in the {\it Drosophila
  Melanogaster}, see SM).
Notably, the best improvement is in the
coherency of the results that the  theory proposed here provides.
The percolation
threshold predicted by the current approximation is always a lower-bound of the
true
percolation threshold, i.e., $p_c \geq p_c^{\textrm{(th)}}$. 
On the contrary, the condition $p_c \geq p_c^{\textrm{(Rad)}}$ is not
granted. 

To summarize, we introduced an exact mathematical framework
able to draw the percolation phase diagram for
arbitrary multiplex networks. We remark that the method describes
the average value of the percolation order parameter over an infinite number of
realizations of the random percolation model. This may not
be representative for specific random realizations 
of the model due to the presence of large fluctuations.
We remark also that the framework relies on the locally treelike ansatz,
so there is still room for
potential corrections to provide better
predictions in loopy multiplexes, such as 
those constructed on the basis of
co-authorship data~\cite{PhysRevE.93.030302}.
Our results obtained from the analysis of real-world
multiplexes confirm the claims of Ref.~\cite{radicchi2015percolation},
in the sense that 
the order parameters predicted by both
theoretical methods exhibit always discontinuous jumps, but their
entity, when one considers the average over 
random disorder, is so small (generally smaller than $10^{-2}$ even on networks
with less than $10^2$ nodes) that they cannot be 
considered as significant. From this perspective, real-world multiplexes 
seem therefore being kept 
cohesive by core edges that do not allow
for abrupt structural transitions.

\begin{acknowledgements}
F.R. acknowledges support from the National Science Foundation
(CMMI-1552487) and the U.S. Army Research Office (W911NF-16-1-0104).
\end{acknowledgements}

%\bibliography{bibliography}

\begin{thebibliography}{37}
\expandafter\ifx\csname natexlab\endcsname\relax\def\natexlab#1{#1}\fi
\expandafter\ifx\csname bibnamefont\endcsname\relax
  \def\bibnamefont#1{#1}\fi
\expandafter\ifx\csname bibfnamefont\endcsname\relax
  \def\bibfnamefont#1{#1}\fi
\expandafter\ifx\csname citenamefont\endcsname\relax
  \def\citenamefont#1{#1}\fi
\expandafter\ifx\csname url\endcsname\relax
  \def\url#1{\texttt{#1}}\fi
\expandafter\ifx\csname urlprefix\endcsname\relax\def\urlprefix{URL }\fi
\providecommand{\bibinfo}[2]{#2}
\providecommand{\eprint}[2][]{\url{#2}}

\bibitem[{\citenamefont{Buldyrev et~al.}(2010)\citenamefont{Buldyrev, Parshani,
  Paul, Stanley, and Havlin}}]{buldyrev2010catastrophic}
\bibinfo{author}{\bibfnamefont{S.~V.} \bibnamefont{Buldyrev}},
  \bibinfo{author}{\bibfnamefont{R.}~\bibnamefont{Parshani}},
  \bibinfo{author}{\bibfnamefont{G.}~\bibnamefont{Paul}},
  \bibinfo{author}{\bibfnamefont{H.~E.} \bibnamefont{Stanley}},
  \bibnamefont{and} \bibinfo{author}{\bibfnamefont{S.}~\bibnamefont{Havlin}},
  \bibinfo{journal}{Nature} \textbf{\bibinfo{volume}{464}},
  \bibinfo{pages}{1025} (\bibinfo{year}{2010}).

\bibitem[{\citenamefont{Boccaletti et~al.}(2014)\citenamefont{Boccaletti,
  Bianconi, Criado, Del~Genio, G{\'o}mez-Garde{\~n}es, Romance,
  Sendi{\~n}a-Nadal, Wang, and Zanin}}]{boccaletti2014structure}
\bibinfo{author}{\bibfnamefont{S.}~\bibnamefont{Boccaletti}},
  \bibinfo{author}{\bibfnamefont{G.}~\bibnamefont{Bianconi}},
  \bibinfo{author}{\bibfnamefont{R.}~\bibnamefont{Criado}},
  \bibinfo{author}{\bibfnamefont{C.~I.} \bibnamefont{Del~Genio}},
  \bibinfo{author}{\bibfnamefont{J.}~\bibnamefont{G{\'o}mez-Garde{\~n}es}},
  \bibinfo{author}{\bibfnamefont{M.}~\bibnamefont{Romance}},
  \bibinfo{author}{\bibfnamefont{I.}~\bibnamefont{Sendi{\~n}a-Nadal}},
  \bibinfo{author}{\bibfnamefont{Z.}~\bibnamefont{Wang}}, \bibnamefont{and}
  \bibinfo{author}{\bibfnamefont{M.}~\bibnamefont{Zanin}},
  \bibinfo{journal}{Physics Reports} \textbf{\bibinfo{volume}{544}},
  \bibinfo{pages}{1} (\bibinfo{year}{2014}).

\bibitem[{\citenamefont{Kivel{\"a} et~al.}(2014)\citenamefont{Kivel{\"a},
  Arenas, Barthelemy, Gleeson, Moreno, and Porter}}]{kivela2014multilayer}
\bibinfo{author}{\bibfnamefont{M.}~\bibnamefont{Kivel{\"a}}},
  \bibinfo{author}{\bibfnamefont{A.}~\bibnamefont{Arenas}},
  \bibinfo{author}{\bibfnamefont{M.}~\bibnamefont{Barthelemy}},
  \bibinfo{author}{\bibfnamefont{J.~P.} \bibnamefont{Gleeson}},
  \bibinfo{author}{\bibfnamefont{Y.}~\bibnamefont{Moreno}}, \bibnamefont{and}
  \bibinfo{author}{\bibfnamefont{M.~A.} \bibnamefont{Porter}},
  \bibinfo{journal}{Journal of complex networks} \textbf{\bibinfo{volume}{2}},
  \bibinfo{pages}{203} (\bibinfo{year}{2014}).

\bibitem[{\citenamefont{Szell et~al.}(2010)\citenamefont{Szell, Lambiotte, and
  Thurner}}]{szell2010multirelational}
\bibinfo{author}{\bibfnamefont{M.}~\bibnamefont{Szell}},
  \bibinfo{author}{\bibfnamefont{R.}~\bibnamefont{Lambiotte}},
  \bibnamefont{and} \bibinfo{author}{\bibfnamefont{S.}~\bibnamefont{Thurner}},
  \bibinfo{journal}{Proceedings of the National Academy of Sciences USA}
  \textbf{\bibinfo{volume}{107}}, \bibinfo{pages}{13636}
  (\bibinfo{year}{2010}).

\bibitem[{\citenamefont{Mucha et~al.}(2010)\citenamefont{Mucha, Richardson,
  Macon, Porter, and Onnela}}]{mucha2010community}
\bibinfo{author}{\bibfnamefont{P.~J.} \bibnamefont{Mucha}},
  \bibinfo{author}{\bibfnamefont{T.}~\bibnamefont{Richardson}},
  \bibinfo{author}{\bibfnamefont{K.}~\bibnamefont{Macon}},
  \bibinfo{author}{\bibfnamefont{M.~A.} \bibnamefont{Porter}},
  \bibnamefont{and} \bibinfo{author}{\bibfnamefont{J.-P.}
  \bibnamefont{Onnela}}, \bibinfo{journal}{science}
  \textbf{\bibinfo{volume}{328}}, \bibinfo{pages}{876} (\bibinfo{year}{2010}).

\bibitem[{\citenamefont{Barth{\'e}lemy}(2011)}]{barthelemy2011spatial}
\bibinfo{author}{\bibfnamefont{M.}~\bibnamefont{Barth{\'e}lemy}},
  \bibinfo{journal}{Physics Reports} \textbf{\bibinfo{volume}{499}},
  \bibinfo{pages}{1} (\bibinfo{year}{2011}).

\bibitem[{\citenamefont{Cardillo et~al.}(2013)\citenamefont{Cardillo,
  G{\'o}mez-Gardenes, Zanin, Romance, Papo, del Pozo, and
  Boccaletti}}]{cardillo2013emergence}
\bibinfo{author}{\bibfnamefont{A.}~\bibnamefont{Cardillo}},
  \bibinfo{author}{\bibfnamefont{J.}~\bibnamefont{G{\'o}mez-Gardenes}},
  \bibinfo{author}{\bibfnamefont{M.}~\bibnamefont{Zanin}},
  \bibinfo{author}{\bibfnamefont{M.}~\bibnamefont{Romance}},
  \bibinfo{author}{\bibfnamefont{D.}~\bibnamefont{Papo}},
  \bibinfo{author}{\bibfnamefont{F.}~\bibnamefont{del Pozo}}, \bibnamefont{and}
  \bibinfo{author}{\bibfnamefont{S.}~\bibnamefont{Boccaletti}},
  \bibinfo{journal}{Scientific Reports} \textbf{\bibinfo{volume}{3}},
  \bibinfo{pages}{1344} (\bibinfo{year}{2013}).

\bibitem[{\citenamefont{Stauffer and Aharony}(1991)}]{stauffer1991introduction}
\bibinfo{author}{\bibfnamefont{D.}~\bibnamefont{Stauffer}} \bibnamefont{and}
  \bibinfo{author}{\bibfnamefont{A.}~\bibnamefont{Aharony}},
  \emph{\bibinfo{title}{Introduction to percolation theory}}
  (\bibinfo{publisher}{Taylor and Francis}, \bibinfo{year}{1991}).

\bibitem[{\citenamefont{Albert et~al.}(2000)\citenamefont{Albert, Jeong, and
  Barab{\'a}si}}]{albert2000error}
\bibinfo{author}{\bibfnamefont{R.}~\bibnamefont{Albert}},
  \bibinfo{author}{\bibfnamefont{H.}~\bibnamefont{Jeong}}, \bibnamefont{and}
  \bibinfo{author}{\bibfnamefont{A.-L.} \bibnamefont{Barab{\'a}si}},
  \bibinfo{journal}{Nature} \textbf{\bibinfo{volume}{406}},
  \bibinfo{pages}{378} (\bibinfo{year}{2000}).

\bibitem[{\citenamefont{Cohen et~al.}(2000)\citenamefont{Cohen, Erez,
  Ben-Avraham, and Havlin}}]{cohen2000resilience}
\bibinfo{author}{\bibfnamefont{R.}~\bibnamefont{Cohen}},
  \bibinfo{author}{\bibfnamefont{K.}~\bibnamefont{Erez}},
  \bibinfo{author}{\bibfnamefont{D.}~\bibnamefont{Ben-Avraham}},
  \bibnamefont{and} \bibinfo{author}{\bibfnamefont{S.}~\bibnamefont{Havlin}},
  \bibinfo{journal}{Phys. {R}ev. {L}ett.} \textbf{\bibinfo{volume}{85}},
  \bibinfo{pages}{4626} (\bibinfo{year}{2000}).

\bibitem[{\citenamefont{Callaway et~al.}(2000)\citenamefont{Callaway, Newman,
  Strogatz, and Watts}}]{callaway2000network}
\bibinfo{author}{\bibfnamefont{D.~S.} \bibnamefont{Callaway}},
  \bibinfo{author}{\bibfnamefont{M.~E.} \bibnamefont{Newman}},
  \bibinfo{author}{\bibfnamefont{S.~H.} \bibnamefont{Strogatz}},
  \bibnamefont{and} \bibinfo{author}{\bibfnamefont{D.~J.} \bibnamefont{Watts}},
  \bibinfo{journal}{Phys. {R}ev. {L}ett.} \textbf{\bibinfo{volume}{85}},
  \bibinfo{pages}{5468} (\bibinfo{year}{2000}).

\bibitem[{\citenamefont{Newman}(2010)}]{newman2010networks}
\bibinfo{author}{\bibfnamefont{M.}~\bibnamefont{Newman}},
  \emph{\bibinfo{title}{Networks: an introduction}} (\bibinfo{publisher}{Oxford
  University Press}, \bibinfo{year}{2010}).

\bibitem[{\citenamefont{Son et~al.}(2012)\citenamefont{Son, Bizhani,
  Christensen, Grassberger, and Paczuski}}]{son2012percolation}
\bibinfo{author}{\bibfnamefont{S.-W.} \bibnamefont{Son}},
  \bibinfo{author}{\bibfnamefont{G.}~\bibnamefont{Bizhani}},
  \bibinfo{author}{\bibfnamefont{C.}~\bibnamefont{Christensen}},
  \bibinfo{author}{\bibfnamefont{P.}~\bibnamefont{Grassberger}},
  \bibnamefont{and} \bibinfo{author}{\bibfnamefont{M.}~\bibnamefont{Paczuski}},
  \bibinfo{journal}{EPL (Europhysics Letters)} \textbf{\bibinfo{volume}{97}},
  \bibinfo{pages}{16006} (\bibinfo{year}{2012}).

\bibitem[{\citenamefont{Baxter et~al.}(2012)\citenamefont{Baxter, Dorogovtsev,
  Goltsev, and Mendes}}]{baxter2012avalanche}
\bibinfo{author}{\bibfnamefont{G.}~\bibnamefont{Baxter}},
  \bibinfo{author}{\bibfnamefont{S.}~\bibnamefont{Dorogovtsev}},
  \bibinfo{author}{\bibfnamefont{A.}~\bibnamefont{Goltsev}}, \bibnamefont{and}
  \bibinfo{author}{\bibfnamefont{J.}~\bibnamefont{Mendes}},
  \bibinfo{journal}{Physical review letters} \textbf{\bibinfo{volume}{109}},
  \bibinfo{pages}{248701} (\bibinfo{year}{2012}).

\bibitem[{\citenamefont{Parshani et~al.}(2010)\citenamefont{Parshani, Buldyrev,
  and Havlin}}]{parshani2010interdependent}
\bibinfo{author}{\bibfnamefont{R.}~\bibnamefont{Parshani}},
  \bibinfo{author}{\bibfnamefont{S.~V.} \bibnamefont{Buldyrev}},
  \bibnamefont{and} \bibinfo{author}{\bibfnamefont{S.}~\bibnamefont{Havlin}},
  \bibinfo{journal}{Physical review letters} \textbf{\bibinfo{volume}{105}},
  \bibinfo{pages}{048701} (\bibinfo{year}{2010}).

\bibitem[{\citenamefont{Min et~al.}(2014)\citenamefont{Min, Do~Yi, Lee, and
  Goh}}]{min2014network}
\bibinfo{author}{\bibfnamefont{B.}~\bibnamefont{Min}},
  \bibinfo{author}{\bibfnamefont{S.}~\bibnamefont{Do~Yi}},
  \bibinfo{author}{\bibfnamefont{K.-M.} \bibnamefont{Lee}}, \bibnamefont{and}
  \bibinfo{author}{\bibfnamefont{K.-I.} \bibnamefont{Goh}},
  \bibinfo{journal}{Physical Review E} \textbf{\bibinfo{volume}{89}},
  \bibinfo{pages}{042811} (\bibinfo{year}{2014}).

\bibitem[{\citenamefont{Bianconi and Dorogovtsev}(2014)}]{bianconi2014multiple}
\bibinfo{author}{\bibfnamefont{G.}~\bibnamefont{Bianconi}} \bibnamefont{and}
  \bibinfo{author}{\bibfnamefont{S.~N.} \bibnamefont{Dorogovtsev}},
  \bibinfo{journal}{Physical Review E} \textbf{\bibinfo{volume}{89}},
  \bibinfo{pages}{062814} (\bibinfo{year}{2014}).

\bibitem[{\citenamefont{Cellai and Bianconi}(2016)}]{cellai2016multiplex}
\bibinfo{author}{\bibfnamefont{D.}~\bibnamefont{Cellai}} \bibnamefont{and}
  \bibinfo{author}{\bibfnamefont{G.}~\bibnamefont{Bianconi}},
  \bibinfo{journal}{Physical Review E} \textbf{\bibinfo{volume}{93}},
  \bibinfo{pages}{032302} (\bibinfo{year}{2016}).

\bibitem[{\citenamefont{Bianconi}(2013)}]{bianconi2013statistical}
\bibinfo{author}{\bibfnamefont{G.}~\bibnamefont{Bianconi}},
  \bibinfo{journal}{Physical Review E} \textbf{\bibinfo{volume}{87}},
  \bibinfo{pages}{062806} (\bibinfo{year}{2013}).

\bibitem[{\citenamefont{Hu et~al.}(2013)\citenamefont{Hu, Zhou, Zhang, Han,
  Rozenblat, and Havlin}}]{hu2013percolation}
\bibinfo{author}{\bibfnamefont{Y.}~\bibnamefont{Hu}},
  \bibinfo{author}{\bibfnamefont{D.}~\bibnamefont{Zhou}},
  \bibinfo{author}{\bibfnamefont{R.}~\bibnamefont{Zhang}},
  \bibinfo{author}{\bibfnamefont{Z.}~\bibnamefont{Han}},
  \bibinfo{author}{\bibfnamefont{C.}~\bibnamefont{Rozenblat}},
  \bibnamefont{and} \bibinfo{author}{\bibfnamefont{S.}~\bibnamefont{Havlin}},
  \bibinfo{journal}{Physical Review E} \textbf{\bibinfo{volume}{88}},
  \bibinfo{pages}{052805} (\bibinfo{year}{2013}).

\bibitem[{\citenamefont{Baxter et~al.}(2016)\citenamefont{Baxter, Bianconi,
  da~Costa, Dorogovtsev, and Mendes}}]{baxter2016correlated}
\bibinfo{author}{\bibfnamefont{G.~J.} \bibnamefont{Baxter}},
  \bibinfo{author}{\bibfnamefont{G.}~\bibnamefont{Bianconi}},
  \bibinfo{author}{\bibfnamefont{R.~A.} \bibnamefont{da~Costa}},
  \bibinfo{author}{\bibfnamefont{S.~N.} \bibnamefont{Dorogovtsev}},
  \bibnamefont{and} \bibinfo{author}{\bibfnamefont{J.~F.}
  \bibnamefont{Mendes}}, \bibinfo{journal}{Physical Review E}
  \textbf{\bibinfo{volume}{94}}, \bibinfo{pages}{012303}
  (\bibinfo{year}{2016}).

\bibitem[{\citenamefont{Cellai et~al.}(2016)\citenamefont{Cellai, Dorogovtsev,
  and Bianconi}}]{PhysRevE.94.032301}
\bibinfo{author}{\bibfnamefont{D.}~\bibnamefont{Cellai}},
  \bibinfo{author}{\bibfnamefont{S.~N.} \bibnamefont{Dorogovtsev}},
  \bibnamefont{and} \bibinfo{author}{\bibfnamefont{G.}~\bibnamefont{Bianconi}},
  \bibinfo{journal}{Physical Review E} \textbf{\bibinfo{volume}{94}},
  \bibinfo{pages}{032301} (\bibinfo{year}{2016}).

\bibitem[{\citenamefont{Radicchi}(2015{\natexlab{a}})}]{radicchi2015percolation}
\bibinfo{author}{\bibfnamefont{F.}~\bibnamefont{Radicchi}},
  \bibinfo{journal}{Nature {P}hys.} \textbf{\bibinfo{volume}{11}},
  \bibinfo{pages}{597} (\bibinfo{year}{2015}{\natexlab{a}}).

\bibitem[{\citenamefont{Cellai et~al.}(2013)\citenamefont{Cellai, L{\'o}pez,
  Zhou, Gleeson, and Bianconi}}]{cellai2013percolation}
\bibinfo{author}{\bibfnamefont{D.}~\bibnamefont{Cellai}},
  \bibinfo{author}{\bibfnamefont{E.}~\bibnamefont{L{\'o}pez}},
  \bibinfo{author}{\bibfnamefont{J.}~\bibnamefont{Zhou}},
  \bibinfo{author}{\bibfnamefont{J.~P.} \bibnamefont{Gleeson}},
  \bibnamefont{and} \bibinfo{author}{\bibfnamefont{G.}~\bibnamefont{Bianconi}},
  \bibinfo{journal}{Physical Review E} \textbf{\bibinfo{volume}{88}},
  \bibinfo{pages}{052811} (\bibinfo{year}{2013}).

\bibitem[{\citenamefont{Min et~al.}(2015)\citenamefont{Min, Lee, Lee, and
  Goh}}]{min2015link}
\bibinfo{author}{\bibfnamefont{B.}~\bibnamefont{Min}},
  \bibinfo{author}{\bibfnamefont{S.}~\bibnamefont{Lee}},
  \bibinfo{author}{\bibfnamefont{K.-M.} \bibnamefont{Lee}}, \bibnamefont{and}
  \bibinfo{author}{\bibfnamefont{K.-I.} \bibnamefont{Goh}},
  \bibinfo{journal}{Chaos, Solitons \& Fractals} \textbf{\bibinfo{volume}{72}},
  \bibinfo{pages}{49} (\bibinfo{year}{2015}).

\bibitem[{\citenamefont{Dorogovtsev et~al.}(2008)\citenamefont{Dorogovtsev,
  Goltsev, and Mendes}}]{dorogovtsev2008critical}
\bibinfo{author}{\bibfnamefont{S.~N.} \bibnamefont{Dorogovtsev}},
  \bibinfo{author}{\bibfnamefont{A.~V.} \bibnamefont{Goltsev}},
  \bibnamefont{and} \bibinfo{author}{\bibfnamefont{J.~F.}
  \bibnamefont{Mendes}}, \bibinfo{journal}{Rev. {M}od. {P}hys.}
  \textbf{\bibinfo{volume}{80}}, \bibinfo{pages}{1275} (\bibinfo{year}{2008}).

\bibitem[{\citenamefont{Radicchi and Castellano}(2016)}]{PhysRevE.93.030302}
\bibinfo{author}{\bibfnamefont{F.}~\bibnamefont{Radicchi}} \bibnamefont{and}
  \bibinfo{author}{\bibfnamefont{C.}~\bibnamefont{Castellano}},
  \bibinfo{journal}{Phys. Rev. E} \textbf{\bibinfo{volume}{93}},
  \bibinfo{pages}{030302} (\bibinfo{year}{2016}).

\bibitem[{\citenamefont{Hamilton and Pryadko}(2014)}]{PhysRevLett.113.208701}
\bibinfo{author}{\bibfnamefont{K.~E.} \bibnamefont{Hamilton}} \bibnamefont{and}
  \bibinfo{author}{\bibfnamefont{L.~P.} \bibnamefont{Pryadko}},
  \bibinfo{journal}{Phys. {R}ev. {L}ett.} \textbf{\bibinfo{volume}{113}},
  \bibinfo{pages}{208701} (\bibinfo{year}{2014}).

\bibitem[{\citenamefont{Karrer et~al.}(2014)\citenamefont{Karrer, Newman, and
  Zdeborov\'a}}]{PhysRevLett.113.208702}
\bibinfo{author}{\bibfnamefont{B.}~\bibnamefont{Karrer}},
  \bibinfo{author}{\bibfnamefont{M.~E.~J.} \bibnamefont{Newman}},
  \bibnamefont{and}
  \bibinfo{author}{\bibfnamefont{L.}~\bibnamefont{Zdeborov\'a}},
  \bibinfo{journal}{Phys. {R}ev. {L}ett.} \textbf{\bibinfo{volume}{113}},
  \bibinfo{pages}{208702} (\bibinfo{year}{2014}).

\bibitem[{\citenamefont{Mezard and Montanari}(2009)}]{mezard2009information}
\bibinfo{author}{\bibfnamefont{M.}~\bibnamefont{Mezard}} \bibnamefont{and}
  \bibinfo{author}{\bibfnamefont{A.}~\bibnamefont{Montanari}},
  \emph{\bibinfo{title}{Information, physics, and computation}}
  (\bibinfo{publisher}{Oxford University Press}, \bibinfo{year}{2009}).

\bibitem[{\citenamefont{Chen et~al.}(2006)\citenamefont{Chen, Hall, and
  Chklovskii}}]{chen2006wiring}
\bibinfo{author}{\bibfnamefont{B.~L.} \bibnamefont{Chen}},
  \bibinfo{author}{\bibfnamefont{D.~H.} \bibnamefont{Hall}}, \bibnamefont{and}
  \bibinfo{author}{\bibfnamefont{D.~B.} \bibnamefont{Chklovskii}},
  \bibinfo{journal}{Proceedings of the National Academy of Sciences of the
  United States of America} \textbf{\bibinfo{volume}{103}},
  \bibinfo{pages}{4723} (\bibinfo{year}{2006}).

\bibitem[{\citenamefont{De~Domenico et~al.}(2014)\citenamefont{De~Domenico,
  Porter, and Arenas}}]{de2014muxviz}
\bibinfo{author}{\bibfnamefont{M.}~\bibnamefont{De~Domenico}},
  \bibinfo{author}{\bibfnamefont{M.~A.} \bibnamefont{Porter}},
  \bibnamefont{and} \bibinfo{author}{\bibfnamefont{A.}~\bibnamefont{Arenas}},
  \bibinfo{journal}{Journal of Complex Networks} p. \bibinfo{pages}{cnu038}
  (\bibinfo{year}{2014}).

\bibitem[{\citenamefont{Stark et~al.}(2006)\citenamefont{Stark, Breitkreutz,
  Reguly, Boucher, Breitkreutz, and Tyers}}]{stark2006biogrid}
\bibinfo{author}{\bibfnamefont{C.}~\bibnamefont{Stark}},
  \bibinfo{author}{\bibfnamefont{B.-J.} \bibnamefont{Breitkreutz}},
  \bibinfo{author}{\bibfnamefont{T.}~\bibnamefont{Reguly}},
  \bibinfo{author}{\bibfnamefont{L.}~\bibnamefont{Boucher}},
  \bibinfo{author}{\bibfnamefont{A.}~\bibnamefont{Breitkreutz}},
  \bibnamefont{and} \bibinfo{author}{\bibfnamefont{M.}~\bibnamefont{Tyers}},
  \bibinfo{journal}{Nucleic acids research} \textbf{\bibinfo{volume}{34}},
  \bibinfo{pages}{D535} (\bibinfo{year}{2006}).

\bibitem[{\citenamefont{De~Domenico
  et~al.}(2015{\natexlab{a}})\citenamefont{De~Domenico, Nicosia, Arenas, and
  Latora}}]{de2015structural}
\bibinfo{author}{\bibfnamefont{M.}~\bibnamefont{De~Domenico}},
  \bibinfo{author}{\bibfnamefont{V.}~\bibnamefont{Nicosia}},
  \bibinfo{author}{\bibfnamefont{A.}~\bibnamefont{Arenas}}, \bibnamefont{and}
  \bibinfo{author}{\bibfnamefont{V.}~\bibnamefont{Latora}},
  \bibinfo{journal}{Nature communications} \textbf{\bibinfo{volume}{6}}
  (\bibinfo{year}{2015}{\natexlab{a}}).

\bibitem[{\citenamefont{De~Domenico
  et~al.}(2015{\natexlab{b}})\citenamefont{De~Domenico, Lancichinetti, Arenas,
  and Rosvall}}]{de2015identifying}
\bibinfo{author}{\bibfnamefont{M.}~\bibnamefont{De~Domenico}},
  \bibinfo{author}{\bibfnamefont{A.}~\bibnamefont{Lancichinetti}},
  \bibinfo{author}{\bibfnamefont{A.}~\bibnamefont{Arenas}}, \bibnamefont{and}
  \bibinfo{author}{\bibfnamefont{M.}~\bibnamefont{Rosvall}},
  \bibinfo{journal}{Physical Review X} \textbf{\bibinfo{volume}{5}},
  \bibinfo{pages}{011027} (\bibinfo{year}{2015}{\natexlab{b}}).

\bibitem[{\citenamefont{Radicchi}(2015{\natexlab{b}})}]{radicchi2014predicting}
\bibinfo{author}{\bibfnamefont{F.}~\bibnamefont{Radicchi}},
  \bibinfo{journal}{Physical Review E} \textbf{\bibinfo{volume}{91}},
  \bibinfo{pages}{010801} (\bibinfo{year}{2015}{\natexlab{b}}).

\bibitem[{\citenamefont{Melnik et~al.}(2011)\citenamefont{Melnik, Hackett,
  Porter, Mucha, and Gleeson}}]{Melnik11}
\bibinfo{author}{\bibfnamefont{S.}~\bibnamefont{Melnik}},
  \bibinfo{author}{\bibfnamefont{A.}~\bibnamefont{Hackett}},
  \bibinfo{author}{\bibfnamefont{M.~A.} \bibnamefont{Porter}},
  \bibinfo{author}{\bibfnamefont{P.~J.} \bibnamefont{Mucha}}, \bibnamefont{and}
  \bibinfo{author}{\bibfnamefont{J.~P.} \bibnamefont{Gleeson}},
  \bibinfo{journal}{Phys. Rev. E} \textbf{\bibinfo{volume}{83}},
  \bibinfo{pages}{036112} (\bibinfo{year}{2011}).

\end{thebibliography}

%\begin{comment}

\clearpage
\newpage
\onecolumngrid

\setcounter{page}{1}
\renewcommand{\theequation}{SM\arabic{equation}}
\setcounter{equation}{0}
\renewcommand{\thefigure}{SM\arabic{figure}}
\setcounter{figure}{0}
\renewcommand{\thetable}{SM\arabic{table}}
\setcounter{table}{0}

\section*{Supplemental Material}

\def\be{\begin{equation}}
\def\ee{\end{equation}}

\def\bc{\begin{center}}
\def\ec{\end{center}}
\def\bea{\begin{eqnarray}}
\def\eea{\end{eqnarray}}

%\gin{CHECKED: Slightly modified}
We consider a multiplex network with $M$ layers and adjacency matrix ${\bf a}^{[\alpha]}$ in each layer $\alpha=1,2,\ldots, M$.
Initially we assume that  we know the set of nodes that are initially damaged. The configuration of the initial damage is indicated by the   variables $\{x_i\}$ where $x_i=0$ ($x_i=1$) if node $i$ is (is not) damaged. 
The message passing  algorithm for given initial damage configuration   determines whether node $i$ belongs ($\sigma_i=1$) or not belongs $\sigma_i=0$ to the mutually connected giant component (MCGC) as long as the multiplex network is locally tree-like.
The algorithm requires the determination of the set of messages 
\bea
\vec{n}_{i\to j}=\left(n_{i \to j}^{[1]},n_{i\to j}^{[2]},\ldots, n^{[\alpha]}_{i\to j}, \ldots,n_{i\to j}^{[M]}\right)
\eea
going from node $i$ to node $j$ connected at least in one layer.
Each message $n_{i\to j}^{[\alpha]}$ indicates whether ($n_{i\to j}^{[\alpha]}=1$) or not ($n_{i\to j}^{[\alpha]}=0$) node $i$ connects node $j$ to the MCGC through links in layer $\alpha$.
These messages are determined by the recursive message passing equations   
\bea
n_{i\to j}^{[\alpha]}=\delta(v_{i\to j},M)a^{[\alpha]}_{ij}x_{i}\left[1-\prod_{\ell\in N(i)\setminus j}\left(1-n^{[\alpha]}_{\ell\to i}\right)\right].
\label{S1}
\eea
Here $v_{i\to j}$ indicates in how many layers node $i$ is connected to the MCGC assuming that node $j$ also belongs to the MCGC and it is given by  
\bea
v_{i\to j}&=&\sum_{\alpha=1}^M\left\{\left[1-\prod_{\ell\in N(i)\setminus j}\left(1-n^{[\alpha]}_{\ell\to i}\right)\right]+a^{[\alpha]}_{ij}\prod_{\ell\in N(i)\setminus j}\left(1-n^{[\alpha]}_{\ell\to i}\right)\right\}.
\label{S2}
\eea
Finally the value of $\sigma_i$ for any generic node $i$ can be expressed in terms of the messages $\vec{n}_{i\to j}$ as
\bea
\hspace{-10mm}\sigma_{i}&=&x_{i}\prod_{\alpha}\left[1-\prod_{\ell \in N(i)}\left(1-n^{[\alpha]}_{\ell\to i}\right)\right].
\label{S3}
\eea
This message passing algorithm  can be applied only  when the full configuration $\{x_i\}$ of the initial damage is known. Here our goal is to derive from this algorithm a distinct  message passing algorithm able to predict the probability $r_i=\Avg{\sigma_i}$ that a node is in the MCGC for a random configuration of the initial damage.
 Specifically we will  assume that the initial damage configuration $\{x_i\}$ has probability 
 \bea
 {\mathcal P}(\{x_i\})=\prod_{i=1}^N p^{x_i}(1-p)^{1-x_i},
 \label{Sps}
 \eea 
 i.e. nodes are independently damaged with probability $f=1-p$.
In order  to predict $r_i$, it is useful to use an alternative formulation of the message passing algorithm for a given configuration of the initial disorder.
This alternative formulation will  allow us to perform easily the average of the initial damage configuration.
To this end, we introduce the variable $\sigma_{i\to j}^{\vec{m},\vec{n}}$ which indicates whether ($\sigma_{i\to j}^{\vec{m},\vec{n}}=1$ ) or not ($\sigma_{i\to j}^{\vec{m},\vec{n}}=0$) node $i$ sends to node $j$  the messages $\vec{n}_{i\to j}$ given that node $i$ and node $j$ are linked by a multilink 
\bea
\vec{m}=\vec{m}_{ij}=\left(a_{ij}^{[1]},a_{ij}^{[2]},\ldots, a_{ij}^{[\alpha]}, \ldots, a_{ij}^{[M]}\right).
\eea 
According to Eqs.(\ref{S1})-(\ref{S2}) a node $i$, in order to send a message $\vec{n}\neq \vec{0}$, should be connected to the MCGC by nodes different from node $j$ in all the layers where $n^{[\alpha]}=1$ and in all the layers where $m^{[\alpha]}=0$. In fact the first requirement is necessary for having $n^{[\alpha]}=1$ the second requirement is necessary for having $v_{i\to j}=M$ because $m^{[\alpha]}=a_{ij}^{[\alpha]}=0$.
Additionally, for every layer $\alpha$ where $m^{[\alpha]}=a_{ij}^{[\alpha]}=1$ but $n^{[\alpha]}=0$ node $i$ must not receive node any positive messages from neighbor nodes different from node $j$.
Therefore we have for  $\vec{n}\neq \vec{0}$,
\bea
\hspace{-10mm}\sigma^{\vec{m},\vec{n}}_{i\to j}&=&x_i\prod_{\alpha=1}^M\left\{\left(m^{[\alpha]}\right)^{n^{[\alpha]}}\left[1-\prod_{\ell\in N(i)\setminus j}\left(1-n^{[\alpha]}_{\ell\to i}\right)\right]^{n^{[\alpha]}m^{[\alpha]}+\left(1-m^{[\alpha]}\right)}\left[\prod_{\ell\in N(i)\setminus j}\left(1-n^{[\alpha]}_{\ell\to i}\right)\right]^{\left(1-n^{[\alpha]}\right)m^{[\alpha]}}\right\},
\label{SF1}
\eea
while for $\vec{n}=\vec{0}$ we have
\bea
\sigma^{\vec{m},\vec{0}}_{i\to j}=1-\sum_{\vec{n}\neq \vec{0}}\sigma^{\vec{m},\vec{n}}_{i\to j}.
\eea
Note that our of the messages $\sigma^{\vec{m},\vec{n}}_{i\to j}$ with different value of $\vec{n}$ only one has value one and all the other are zero. We call this message $\vec{n}_{i\to j}$ or in other words, 
\bea
\vec{n}_{i\to j}=\mbox{argmax}_{\vec{n}}\sigma^{\vec{m},\vec{n}}_{i\to j}.
\eea
This different formulation of the message passing equations, is suitable to easy perform an average that takes into account the correlations existing between the different messages $n_{i\to j}^{[\alpha]}$ between node $i$ and node $j$.
In order to perform the average over the probability ${\mathcal P}(\{x_i\})$ given by Eqs. $(\ref{Sps})$, let us use the identity valid for $p^{[\alpha]}$ taking values $p^{[\alpha]}=0,1$
\bea
	&&\prod_{\alpha=1}^M (1-x_{\alpha})^{p^{[\alpha]}}=\prod_{\alpha|p^{[\alpha]}>0}(1-z_{\alpha}) = \sum_{\vec{r}|r^{[\alpha]}=0 \ \mbox{{\scriptsize if}}\  p^{[\alpha]}=0}(-1)^{\sum_{\alpha=1}^M r^{[\alpha]}} \left(z_{\alpha}\right)^{r^{[\alpha]}},
	\label{r}
\eea
where the sum in the last term is over all the vectors 
\bea
\vec{r}=\left(r^{[1]},r^{[2]},\ldots, r^{[\alpha]},\ldots, r^{[M]}\right)
\eea
of elements $r^{[\alpha]}=0,1$ for $p^{[\alpha]}=1$ and $r^{[\alpha]}=0$ for $p^{[\alpha]}=0$.
Using this relation for Eq. $(\ref{SF1})$ we obtain 
\bea
\hspace*{-13mm}\sigma^{\vec{m},\vec{n}}_{i\to j}= x_i\sum_{\vec{r}|r^{[\alpha]}=0 \ \mbox{if}\ (1-n^{[\alpha]})m^{[\alpha]}=1}\left[\prod_{\alpha=1}^M \left(m^{[\alpha]}\right)^{n^{[\alpha]}}\right](-1)^{\sum_{\alpha}r^{[\alpha]}}\prod_{\ell\in N(i)\setminus j}\prod_{\alpha=1}^M\left(1-n_{\ell\to i}^{[\alpha]}\right)^{r^{[\alpha]}+m^{[\alpha]}\left(1-n^{[\alpha]}\right)}.
\eea
Since    between all the messages $\sigma_{i\to j}^{\vec{m},\vec{n}}$ sent between node $i$ to node $j$ only one message is equal to one, we have 
\bea
\hspace*{-13mm}\sigma^{\vec{m},\vec{n}}_{i\to j}= x_i\sum_{\vec{r}|r^{[\alpha]}=0 \ \mbox{if}\ (1-n^{[\alpha]})m^{[\alpha]}=1}\left[\prod_{\alpha=1}^M \left(m^{[\alpha]}\right)^{n^{[\alpha]}}\right](-1)^{\sum_{\alpha}r^{[\alpha]}}\prod_{\ell\in N(i)\setminus j}\left(1-\sum_{\vec{n}'|\sum_{\alpha}\left(n^{\prime}\right)^{{[\alpha]}}[r^{[\alpha]}+(1-n^{[\alpha]})m^{[\alpha]}]>0}\sigma_{\ell\to i}^{\vec{m}_{\ell i}\vec{n}'}\right).
\eea
By averaging these messages over the distribution ${\mathcal P}(\{x_i\})$ given by Eq. (\ref{SF1}) we can formulate a different message passing algorithm able to predict the probability $r_i$ that a random node belongs to the MCGC for a random realization of the initial disorder. In this case the generic message $s^{\vec{m}_{ij},\vec{n}}_{i\to j}$ indicates the probability that node $i$ connects node $j$ to the MCGC in the layers where ${n}^{[\alpha]}=1$.
These messages are given by $s^{\vec{m}_{ij},\vec{n}}_{i\to j}=\Avg{\sigma^{\vec{m},\vec{n}}_{i\to j}}$ where the average is over the random realization of the initial disorder. Therefore they satisfy the following recursive equations 
\bea
\hspace*{-13mm}s^{\vec{m},\vec{n}}_{i\to j}= p\sum_{\vec{r}|r^{[\alpha]}=0 \ \mbox{if}\ (1-n^{[\alpha]})m^{[\alpha]}=1}\left[\prod_{\alpha=1}^M \left(m^{[\alpha]}\right)^{n^{[\alpha]}}\right](-1)^{\sum_{\alpha}r^{[\alpha]}}\prod_{\ell\in N(i)\setminus j}\left(1-\sum_{\vec{n}'|\sum_{\alpha}\left(n^{\prime}\right)^{{[\alpha]}}[r^{[\alpha]}+(1-n^{[\alpha]})m^{[\alpha]}]>0}s_{\ell\to i}^{\vec{m}_{\ell i}\vec{n}'}\right),
\eea
as long as the multiplex network is locally tree-like.
Similarly  the probability $r_i$ that node $i$ is in the MCGC is the average $r_i=\Avg{\sigma_i}$, i.e.  
\bea
r_i=p\sum_{\vec{r}}(-1)^{\sum_{\alpha}r^{[\alpha]}}\left[\prod_{\ell\in N(i)}\left(1-\sum_{\vec{n}'|\sum_{\alpha}\left(n^{\prime}\right)^{[\alpha]}r^{[\alpha]}>0}s_{\ell\to i}^{\vec{m}_{\ell i}\vec{n}'}\right)\right],
\eea
as long as the multiplex network satisfy the locally tree-like approximation.

%%%%%%%%%%%
\begin{figure*}[!htb]
\includegraphics[width=0.9\textwidth]{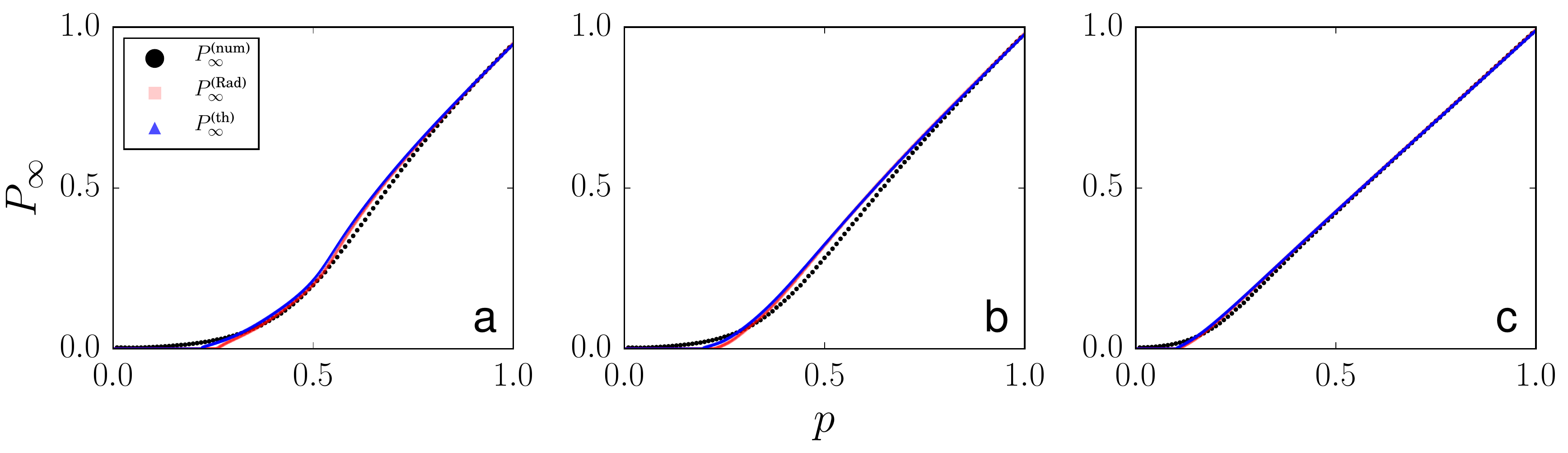}
\caption{Percolation diagrams for the {\it Caenorhabditis Elegans} duplex
  networks. Description of the various panels are identical to those
  of Fig.~\ref{figure} of the main text. Order of appearance the duplexes is
  identical to the one of Table~\ref{table} of the main text.
}
\label{figure_ce}
\end{figure*}
%%%%%%%%%%%

%%%%%%%%%%%
\begin{figure*}[!htb]
\includegraphics[width=0.9\textwidth]{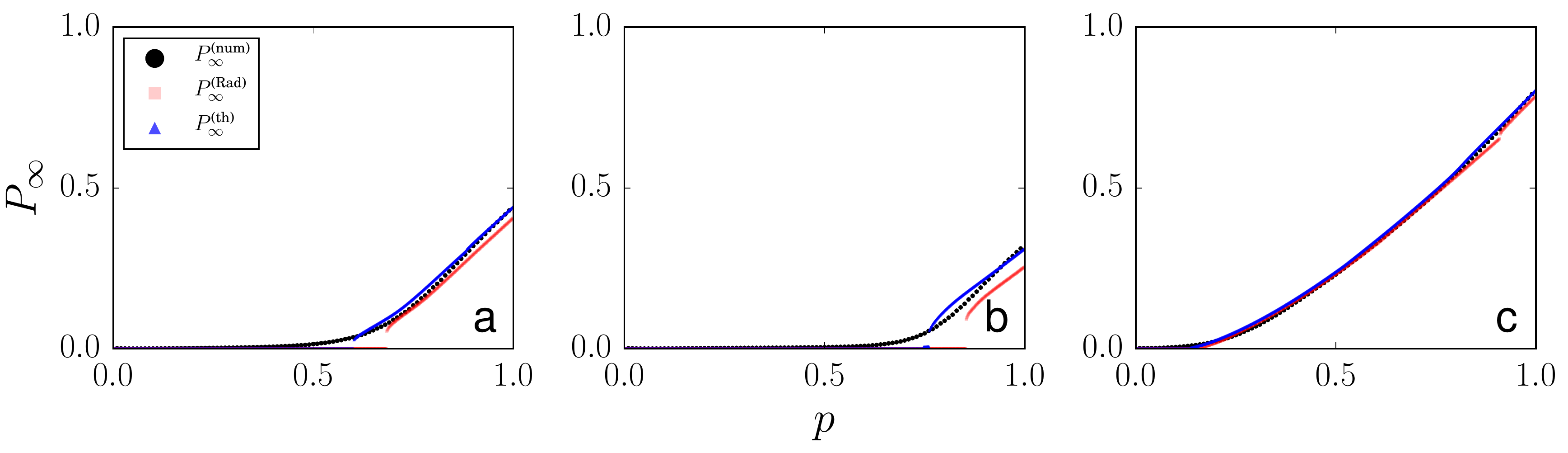}
\caption{Percolation diagrams for the {\it Drosophila
                Melanogaster} duplex
  networks. Description of the various panels are identical to those
  of Fig.~\ref{figure} of the main text. Order of appearance the duplexes is
  identical to the one of Table~\ref{table} of the main text.
}
\label{figure_dro}
\end{figure*}
%%%%%%%%%%%

%%%%%%%%%%%
\begin{figure*}[!htb]
\includegraphics[width=0.9\textwidth]{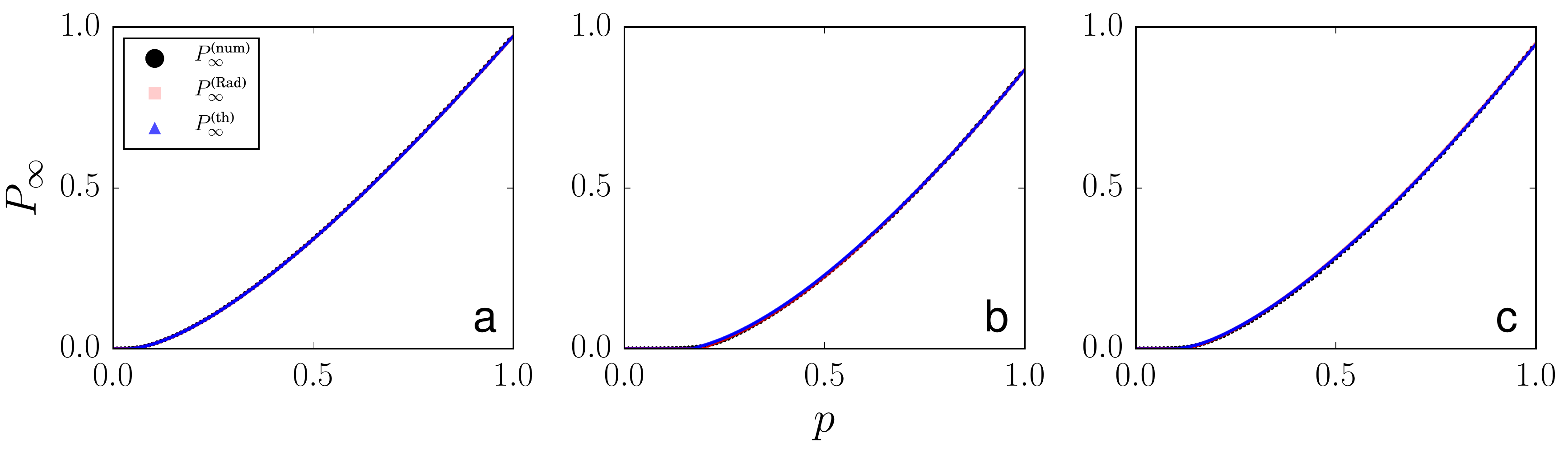}
\caption{Percolation diagrams for the {\it Homo Sapiens} duplex
  networks. Description of the various panels are identical to those
  of Fig.~\ref{figure} of the main text. Order of appearance the duplexes is
  identical to the one of Table~\ref{table} of the main text.
}
\label{figure_homo}
\end{figure*}
%%%%%%%%%%%

%%%%%%%%%%%
\begin{figure*}[!htb]
\includegraphics[width=0.9\textwidth]{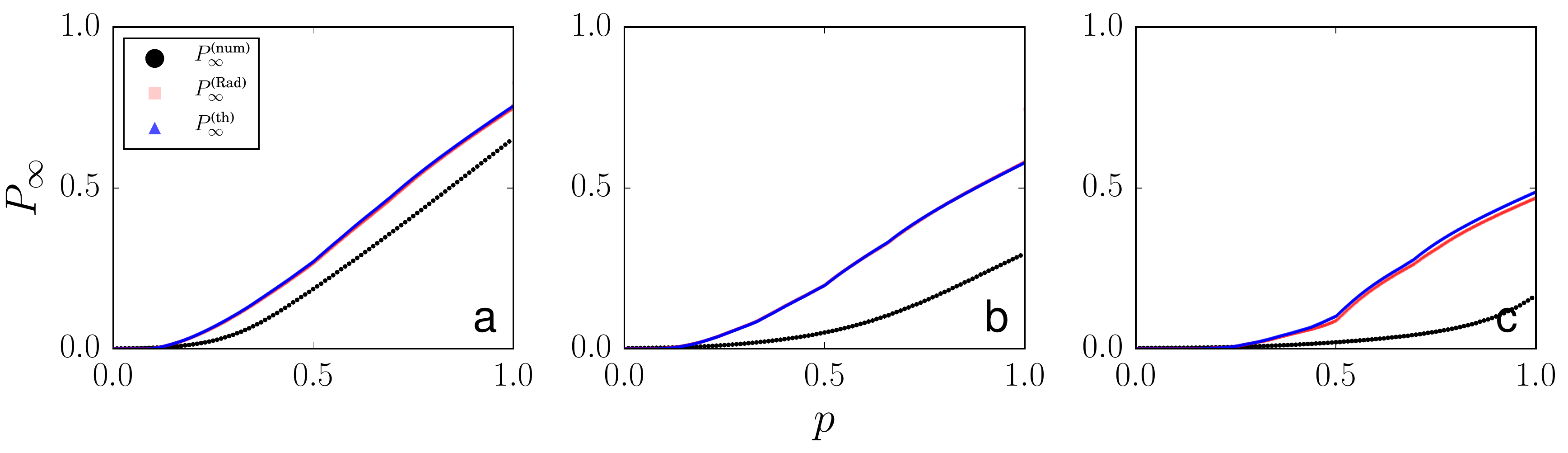}
\caption{Percolation diagrams for the {\it NetSci Co-authorship} duplex
  networks. Description of the various panels are identical to those
  of Fig.~\ref{figure} of the main text. Order of appearance the duplexes is
  identical to the one of Table~\ref{table} of the main text.
}
\label{figure_arxiv}
\end{figure*}
%%%%%%%%%%%

%\end{comment}

\end{document}